\documentclass[preprint,showpacs,preprintnumbers,amsmath,amssymb]{revtex4}

\usepackage{graphicx}
\usepackage{dcolumn}
\usepackage{bm}

\begin{document}

\title{Capillary and anchoring effects in thin hybrid nematic films and connection with
bulk behaviour}

\author{D. de las Heras}
\affiliation{Departamento de F\'{\i}sica Te\'orica de la Materia Condensada,
Universidad Aut\'onoma de Madrid, E-28049 Madrid, Spain}

\author{Luis Mederos}
\affiliation{Instituto de Ciencia de Materiales de Madrid, Consejo Superior de Investigaciones Cient\'{\i}ficas, Sor Juana In\'es de la Cruz, 3, E-28049 Madrid, Spain}

\author{Enrique Velasco}
\affiliation{Departamento de F\'{\i}sica T\'{e}orica de la Materia Condensada
and Instituto de Ciencia de Materiales Nicol\'as Cabrera,
 Universidad Aut\'{o}noma de Madrid, E-28049 Madrid, Spain}

\date{\today}

\begin{abstract}
By means of a molecular model, we examine hybrid nematic films with 
antagonistic anchoring angles where one of the surfaces is in the
strong anchoring regime. If anchoring at the other surface is weak, 
and in the absence of wetting by the isotropic
phase, the anchoring transition may interact with the capillary
isotropic--nematic transition in interesting ways. For general
anchoring conditions on this surface we confirm the existence of the step--tilt, 
biaxial phase and the associated transition to the linear, 
constant--tilt--rotation, configuration. The step--like phase is 
connected with the bulk isotropic phase for increasing film thickness so that
the latter transition is to be interpreted as the capillary isotropic--nematic
transition. Finally, we suggest possible global surface phase diagrams. 
\end{abstract}


\maketitle
\begin{figure}[h]
\includegraphics[scale=0.7]{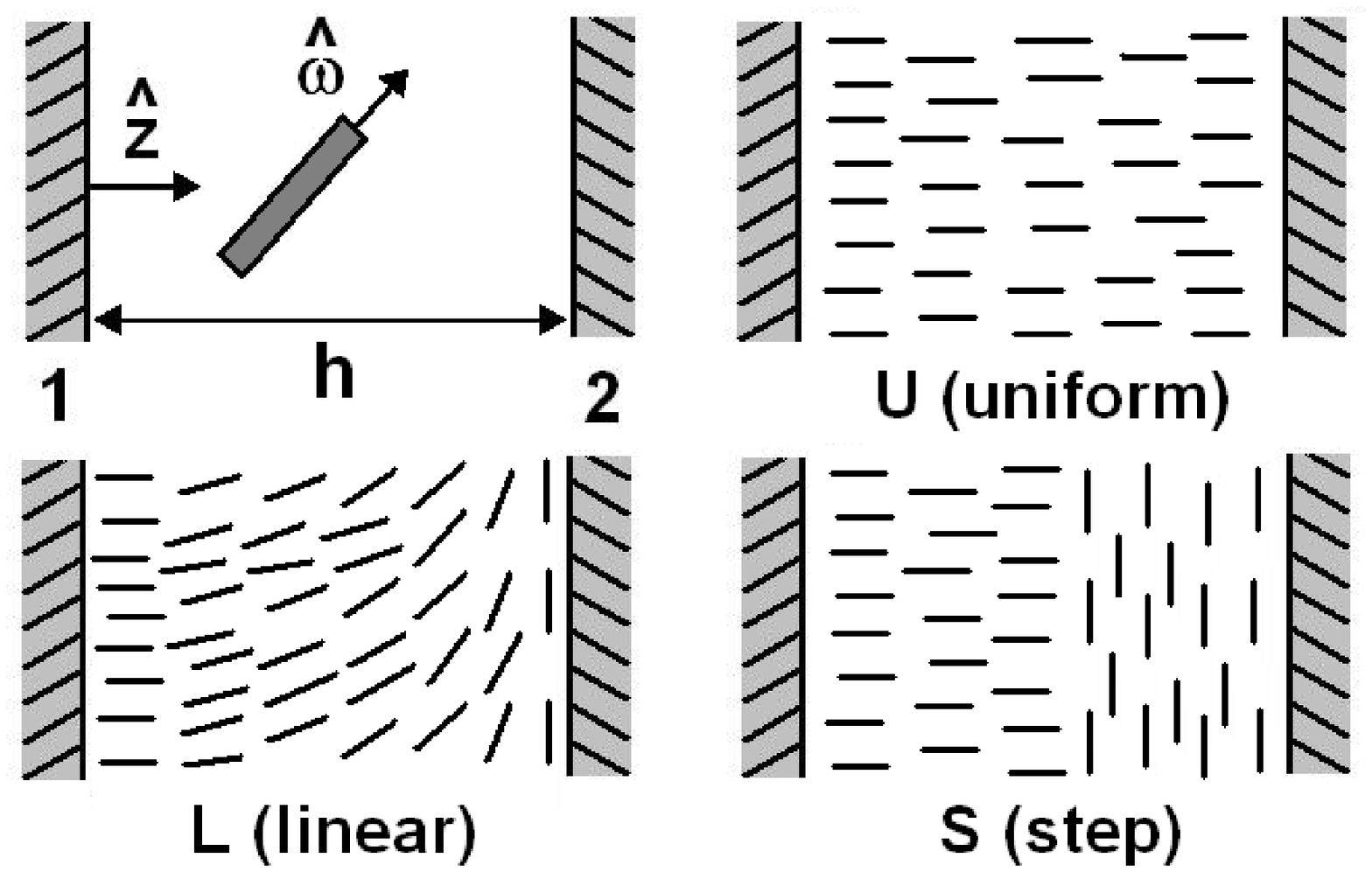}
\caption{Geometry used for the calculations and director
configurations for the three possible phases inside a hybrid
cell. The U phase, here represented with
homeotropic orientation, may also have planar orientation,
depending on the relative anchoring energies of the two
antagonistic substrates.}
\label{fig1}
\end{figure}

Frustration effects associated with confinement of liquid crystals
by competing surface fields continue to attract interest.
Hybrid nematic cells, where a nematic material is exposed to two surfaces with
strong but opposing anchoring tendencies, are commonly used experimentally to 
determine 
anchoring properties of substrates \cite{Blinov,Mazzulla} and may be important
in the context of display device technologies, especially in the case of
hybrid twist nematic cells \cite{Virga}.
The two configurations observed in hybrid hometropic/planar
cells are (Fig. \ref{fig1}): (i) uniform
director along the direction favoured by the substrate with the highest anchoring
energy (U phase), and (ii) linearly--rotating director between the two antagonistic surfaces (L phase). 
Elastic and surface effects play opposite roles
and determine the equilibrium director configuration. 
A confinement--induced anchoring transition (between the L and U phases)
is expected for sufficiently thin cells \cite{Barbero}.
However, a third possible
configuration was proposed by Palffy--Muhoray et al. \cite{Palffy1994}
(and a few years before by Schopohl and Sluckin \cite{Schopohl} in the
context of disclination defects), consisting
of two contiguous slabs of nematic material with uniform but opposite director
orientations, each following the orientation favoured by each substrate; 
macroscopically the directors in the two slabs cannot be
continuously connected. This phase will be called {\it step--like} (S) phase 
(see Fig. \ref{fig1}), but is also known as director--exchange phase,
biaxial phase, etc.
Although the S phase, predicted by the phenomenological Landau--de Gennes 
theory \cite{Palffy1994,Galabova,Sarlah}, 
has been confirmed by Monte Carlo simulation on a (lattice--spin) 
Lebwohl--Lasher model \cite{Chiccoli}, 
no experimental evidence seems to exist as yet (see however Ref. \cite{Zap}),
presumably because 
the step--to--linear (SL) phase transition will take place very close to the
bulk isotropic--nematic transition and for a cell thickness of a few tens of
nanometers \cite{Sarlah}, challenging experimental verification. Also, an
understanding of the role of the different phases in the surface phase 
diagram and of
the relation of the capillary isotropic--nematic transition in hybrid cells
to bulk behaviour (in particular, to the anchoring transition for a single 
substrate) is, we believe, still lacking.

In this Letter we establish the connection between capillary and anchoring
effects, on the one hand, and clarify the role of the SL transition in the surface
phase diagram, on the other, in an asymmetric
nematic film confined between dissimilar, parallel substrates. 
In addition, we discuss possible scenarios, in the weak
anchoring regime, for how this connection may be realised. 
In contrast with previous theoretical works based on mesoscopic models, 
the problem 
is analysed using a mean--field molecular theory 
applied to an athermal fluid of hard anisotropic particles that undergoes 
an isotropic--nematic transition with respect to chemical potential
(which plays the role of an inverse temperature in a thermotropic liquid crystal).
The model, extensively tested previously \cite{Dani1,Dani2},
consistently describes bulk and interfacial properties, allowing
for a microscopically--based assessment of the effects of surface 
interactions, wetting properties and elastic free energies, on the formation of
the S phase. Our analysis indicates that the SL transition corresponds in fact to the usual
capillary isotropic--nematic transition, well described, in the limit of thick planar cells,
by the Kelvin equation, which relates the transition to the wetting conditions of
the substrates. In addition, we relate the capillary transition to the anchoring
transition in the confined system when one of the substrates is in the
weak anchoring regime, and provide several possible scenarios for the global
surface phase diagram.

\begin{figure}
\includegraphics[scale=0.95]{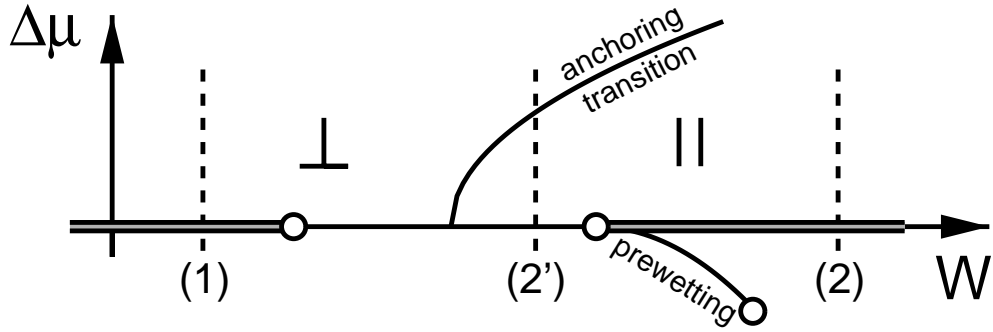}
\caption{Interfacial phase diagram of the single--substrate model. 
$\Delta\mu$ is the chemical potential relative
to bulk coexistence and $W$ the surface strength. $\perp$ and $\parallel$ 
indicate nematic bulk director configurations perpendicular (homeotropic) and 
parallel to substrate, respectively. Thick lines are wetting transitions.
Thin lines are anchoring and prewetting transitions for substrate 2.
(1) is the state chosen for substrate 1 (complete wetting and strong anchoring) 
whereas (2) and (2') are two states for substrate 2 considered in the text.}
\label{fig2}
\end{figure}

The theoretical model is the density--functional theory of 
Onsager, with Parsons--Lee rescaling (see \cite{Dani1,Dani2} for a full
account of the theory and its numerical implementation), formulated for rigid particles 
(spherocylinders) with a length--to--breadth ratio $L/D=5$. In this theory
the free energy $F[\rho]$ is minimised with respect to the density
distribution, $\rho(z,\hat{\bf\omega})$, with $z$ the normal distance from
the left substrate and $\hat{\bf\omega}$ the main axis of the uniaxial
rods (Fig. \ref{fig1}). The theory includes
interactions from excluded volume effects, which promote nematic ordering at high 
packing fractions, and entropic contributions from orientational degrees
of freedom, which play against those effects. A bulk isotropic--nematic transition
is predicted for sufficiently slender rods at a chemical potential $\mu_{\rm b}$.
Four structural quantities are needed to describe ordering: 
$\rho(z)$, the local number density, $\eta(z)$ and $\sigma(z)$,
the uniaxial and biaxial nematic order parameters with respect to the local 
director $\hat{\bf n}$, and $\psi(z)$, the tilt angle of the local director 
with respect to the substrate 
normal along the unit vector $\hat{\bf z}$. Further, the effect of each substrate 
is accounted for by means of a one--particle external potential, which for the 
substrate at left is $V_{\rm ext}^{(1)}(z,\hat{\bf\omega})=\infty$ ($z<0$) and 
$W_{1}\exp{\left(-\alpha z\right)} P_2(\hat{\bf\omega}\cdot\hat{\bf z})$ ($z>0$),
with a similar expression for the substrate potential 
$V_{\rm ext}^{(2)}(z,\hat{\bf\omega})$ at right (but with a strength $W_2$).
In all our calculations we set the potential range as $\alpha=0.88(L+D)^{-1}$.
This model has been used \cite{Dani1,Dani2} to explore wetting properties; 
it also predicts an 
anchoring transition \cite{Dani1}. Fig. \ref{fig2} is the (schematic) interfacial 
phase diagram for a single--substrate system in the 
$\Delta\mu\equiv\mu-\mu_{\rm b}$ vs. surface strength $W$ plane.
Dashed vertical lines indicate the state of both substrates in the confined case; substrate 1 is
fixed to be in the regime of complete wetting by nematic (with homeotropic
orientation $\perp$) and in the strong anchoring case, while the state of substrate 2 will be 
chosen in the regimes of complete wetting (with planar orientation $\parallel$) or
partial wetting. In the latter case proximity to the anchoring transition
will bring about interaction of anchoring and capillary effects in the 
confined system. The model does not predict a region of wetting 
by the isotropic phase; consequently, we expect no suppression of the capillary
isotropic--nematic transition \cite{Quintana,Inma1} in our system.

\begin{figure}[h]
\includegraphics[scale=0.30]{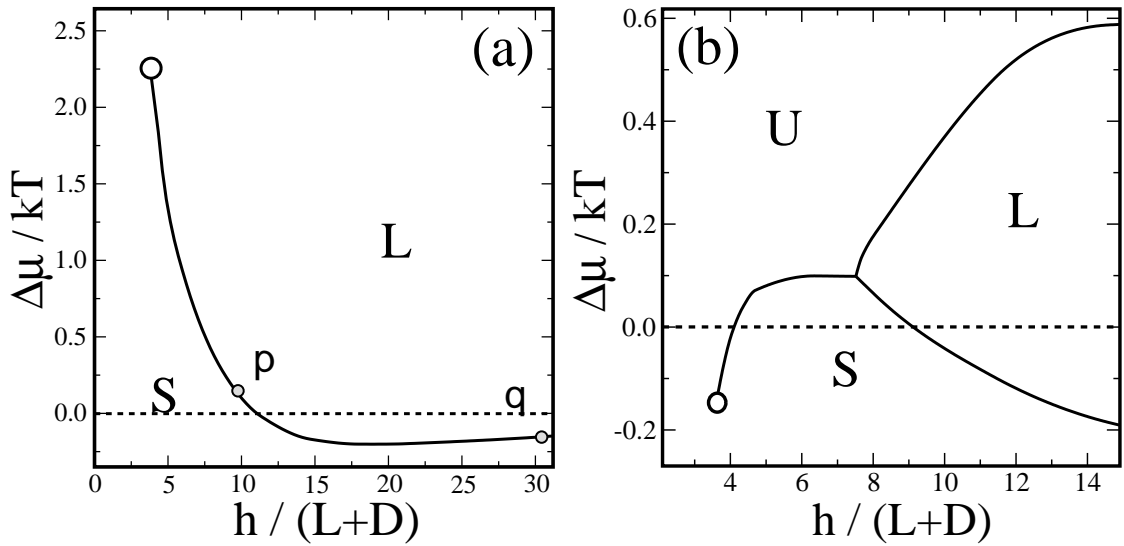}
\caption{Interfacial phase diagram in the chemical potential
$\Delta\mu$ vs. pore width $h$ plane, for values of the surface
strengths (a) $W_1=-1kT$, $W_2=3kT$, and (b) $W_1=0$, $W_2=0.35kT$. 
Points p and q in (a) are the two coexistence points for which order--parameter 
and tilt--angle profiles are shown in Fig. \ref{fig4}, as discussed in the text. 
Open circles are surface critical point.}
\label{fig3}
\end{figure}

Our first result is Fig. \ref{fig3}(a), which depicts the surface phase diagram 
in the $\Delta\mu$ vs.  film thickness $h$ plane. 
Conditions of wetting by nematic and strong anchoring with homeotropic and planar 
alignments [states (1) and (2) in Fig. \ref{fig2}] are imposed; therefore no anchoring 
transition is expected in the nematic region. The line represents a first--order SL phase transition
separating two confined structures, the L and S phases. For thin films the line terminates 
in a critical point. Therefore, the stable 
director configuration for sufficiently thin films is the S phase
(as predicted by Palffy--Muhoray et al. \cite{Palffy1994}), but only
below the critical
point (provided the latter is within the stable nematic region). 
As $h\to\infty$ (thick films) the transition line 
approches the bulk value $\mu_b$ from below, as corresponds to a preference of
the substrate for the nematic phase. 
The non--monotonic shape of the transition line can be understood from a competition
between capillary and elastic effects in the L phase: macroscopically 
$\Delta\mu(h)=a_2/h^2-a_1/h$, where $a_2>0$ is proportional to the nematic
elastic constant $K$ and $a_1>0$ to the surface tension difference $\Delta\gamma=2\gamma_{\rm WI}-
\gamma_{\rm WN}^{\parallel}-\gamma_{\rm WN}^{\perp}$ (the last two 
surface tensions corresponding to planar and homeotropic director
configurations, respectively); as $h\to\infty$ capillary effects [second term
in $\Delta\mu(h)$] dominate and $\Delta\mu(h)\to 0^{-}$, but in the thin--film regime 
elastic effects are more important and $\Delta\mu$ becomes positive. 

\begin{figure}
\includegraphics[scale=0.48]{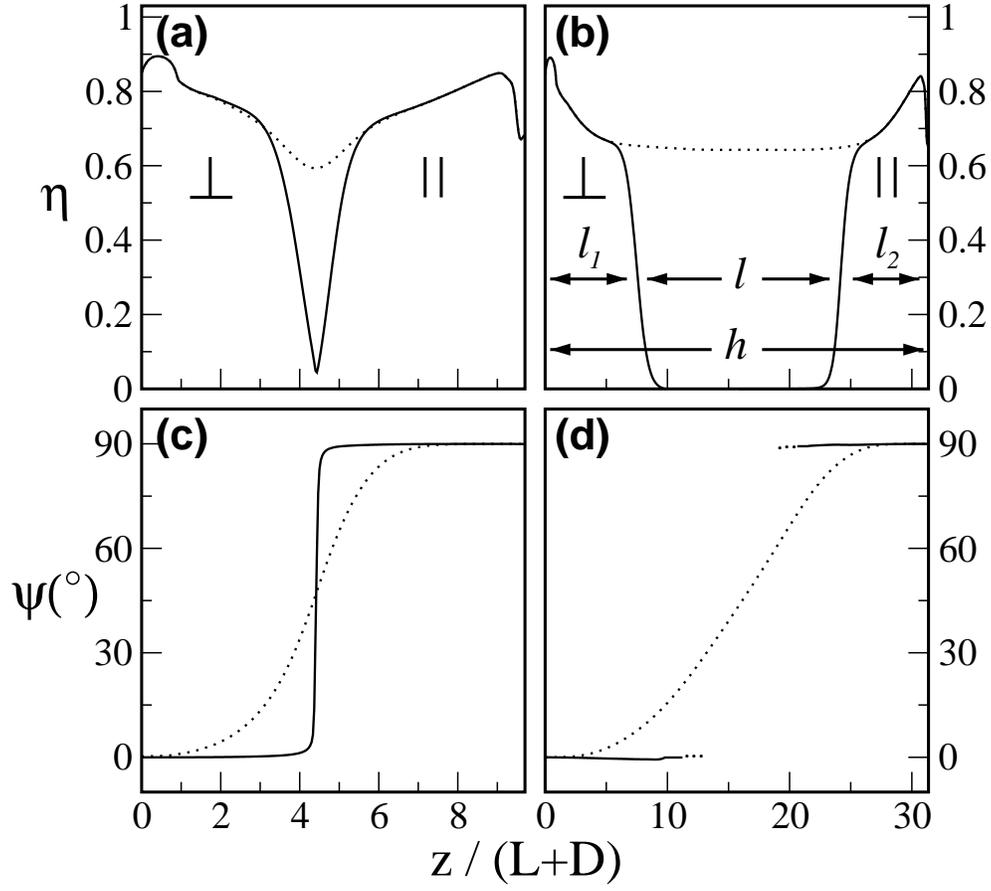}
\caption{Uniaxial order--parameter $\eta(z)$ and tilt--angle $\psi(z)$
profiles for the two film structures that coexist at points
p (a and c) and q (b and d) in the phase diagram of Fig. \ref{fig3}(a). 
Continuous line: S phase; dotted line: L phase. In (d) the tilt
angle profile in the isotropic region is not plotted since it cannot
be defined.}
\label{fig4}
\end{figure}

What is the structure of the L and S phases? This is shown in
Fig. \ref{fig4}, where the order parameter profiles
of the two structures coexisting at point p (a and c) and point q
(b and d) on the transition line depicted in Fig. \ref{fig3}(a) are 
shown. Particularly interesting
is the behaviour of $\eta$ in the S phase in the central region of the cell; 
it drops almost to zero symmetrically with respect to the step, 
Fig. \ref{fig4}(a). One can interpret this profile as indicating the presence of
a planar defective region sandwiched 
between two finite--thickness nematic films with opposite director orientations,
i.e. a `true' S phase. However, in the S--type structure of point q
[Fig. \ref{fig4}(b)], this region has evolved into a well--developed
isotropic slab whose thickness $l$ depends on the 
departure from the bulk, $\Delta\mu$: if $l_1$ and $l_2$ are the thicknesses of the incipent 
wetting nematic layers, then [see Fig. \ref{fig4}(b)]
$l=h-l_1-l_2\sim -\Delta\mu^{-1}$ according to Kelvin equation (large $h$);
it follows that $l$ is an
increasing function of $h$ as $h\to\infty$ for the coexisting S phase and, 
for large values of $h$, the isotropic slab completely decouples the two
nematic films with opposite director orientations \cite{elastic}, 
the situation being
identical to the usual capillary isotropic--nematic transition in 
symmetric or nearly symmetric cells. Therefore, the S phase is in fact
a confined phase connected with the bulk isotropic (I) 
phase, whereas the L phase corresponds to the confined phase connected with the
bulk nematic (N) phase: we are observing the usual capillary IN transition
line.

The S phase, therefore, is not a genuine phase different from
the confined I phase: whether the two nematic films are in contact or not will
depend on conditions such as wetting strength and departure
from bulk coexistence (determining film thickness) of the particular
material/surface. What we can say is that the optimum conditions to observe 
a `true' S phase, with an intervening defective region of {\it molecular} thickness 
where the tilt angle changes abruptly (see Fig. \ref{fig1}), involve nematic wetting
at both substrates, sufficiently narrow pore widths and sufficiently 
developed nematic layers adsorbed at the two substrates (or equivalently,
closeness to the bulk phase transition). Both conditions may play against 
univocal experimental verification. These observations are implicit in
the recent paper by Chiccoli et al. \cite{Chiccoli} who performed Monte 
Carlo simulations on a (lattice) Lebwohl--Lasher model. However, the
authors established a maximum cell thickness in which the S phase
could be found. Our results rather suggest that: (i) the S phase can
be found for any value of cell thickness, (ii) the SL transition continues
to bulk as the true capillary transition, and (iii) wetting properties, 
along with closeness to bulk transition and pore width, determine 
whether a `true' step--like phase can be observed or not.

\begin{figure}
\includegraphics[scale=0.65]{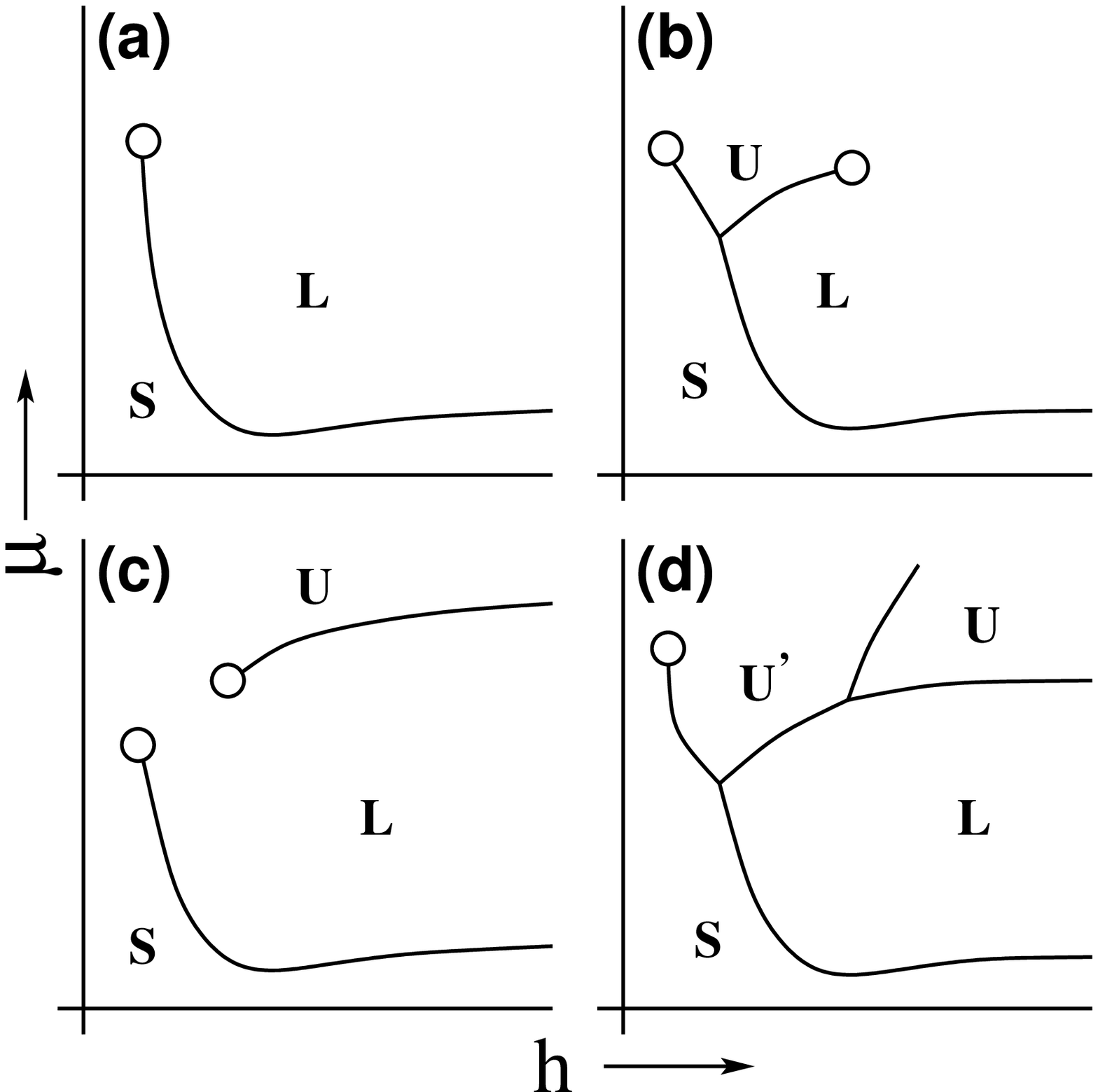}
\caption{Possible phase diagrams for confined films in hybrid cell (schematic).
(a) Single SL transition with strong anchoring conditions
on both surfaces. (b) Additional U phase
and associated UL (confinement--induced) anchoring transition when
one anchoring energy weaker than the other.
(c) Additional anchoring transition 
when corresponding substrate is in weak anchoring regime. (d) Two 
anchoring transition lines, UL and U$^{\prime}$L,
the first continued from the bulk and the second induced by
confinement when at least one substrate is 
in weak anchoring regime. Fig. \ref{fig3}(b)
pertains to case (c) where the two transition lines meet at a triple point.}
\label{fig5}
\end{figure}

The central result of our analysis is that the SL transition line may 
in some cases interact with the anchoring 
phase transition occurring in the semiinfinite case, in the weak anchoring 
regime, when the chemical potential is varied [state (2') in Fig. \ref{fig2}]. 
In our model, the anchoring transition in substrate 2 becomes a UL phase transition 
in the confined system, but this transition is genuinely 
different from the SL transition; the two can be present at the same time
and may in fact interact in some 
cases in the regime of very narrow pores. This is illustrated 
in Fig. \ref{fig3}(b), which shows the phase diagram for a case
pertaining to complete and partial wetting by nematic, respectively, but with an anchoring transition
occurring in the latter substrate. In this case a ULS triple point 
actually occurs. However, other scenarios, where e.g. the SL and UL lines 
do not meet, may be possible. Also, in the strong anchoring regime,
where no anchoring transition exists in the semiinfinite system, 
a confinement--induced anchoring (UL) transition could exist for narrow pores 
\cite{Barbero}; Fig. \ref{fig5}
presents a summary of possible scenarios that can actually be obtained
by our theoretical model. 

Rodr\'{\i}guez--Ponce et al. \cite{Inma,Inma1} have analysed a related system using 
a simplified version of density--functional theory. However, their system is
crucially different in that the isotropic phase wets the substrate that
undergoes the anchoring transition; the result is that the capillary isotropic--nematic
phase transition is suppressed, and capillary and anchoring transitions never occur 
at the same time in the weak anchoring regime: no interaction between the two is possible. 
For the same reason, the U phase is never stabilised and the transition in the
confined system always proceeds between the S and L phases.
Suppression of the capillary transition is expected whenever 
the isotropic phase preferentially adsorbs 
on (or wets) one of the substrates \cite{Inma1} (so that one of the nematic 
phases, either hometropic or planar, becomes irrelevant) or the director 
orientation of the planar nematic film is random \cite{Quintana}, the
phenomenology being similar to that in magnetic systems \cite{Ising} (with
some variations related to elastic effects).
 
In summary, we have presented a scenario, based on a microscopic model,
for the phase equilibria of a liquid--crystal film subject to opposite
anchoring energies in a planar cell. The different possible director
structures, their phase boundaries and their relation
to the bulk and anchoring transitions of the
corresponding system adsorbed on a single substrate, have been discussed.
We believe our work clarifies recent analyses on the hybrid cell,
based on phenomenological approaches, Monte Carlo simulation on a
lattice spin model, and density--functional theory.
Therefore our work presents a unified picture of the capillary phenomena 
occurring in a hybrid cell, linking phenomena such as the SL transition
to capillary nematization, wetting and anchoring. 

We acknowledge financial support from Ministerio de Educaci\'on y
Ciencia (Spain) under grants Nos. FIS2005-05243-C01-01 and
FIS2007-65869-C03-01, and Comunidad Aut\'onoma de Madrid
(Spain) under grant No. S-0505/ESP-0299.

\newpage

\newpage

\newpage

\newpage

\newpage

\begin{thebibliography}{13}
\bibitem{Blinov} L. M. Blinov, D. B. Subachyus, and S. V. Yablonsky,
J. Phys. II {\bf 1}, 459 (1991).
\bibitem{Mazzulla} A. Mazzulla, F. Ciuchi, and J. R. Sambles, Phys. Rev. E {\bf 64},
021708 (2001).
\bibitem{Virga} F. Bisi, E. C. Gartland, R. Rosso, and E. G. Virga,
Phys. Rev. E {\bf 68}, 0217078 (2003).
\bibitem{Barbero} G. Barbero and R. Barberi, J. Phys. (Paris) {\bf 44}, 609 (1983).
\bibitem{Palffy1994} P. Palffy--Muhoray, E. C. Gartland, and J. R. Kelly,
Liq. Crys. {\bf 16}, 713 (1994).
\bibitem{Schopohl} N. Schopohl and T. J. Sluckin, Phys. Rev. Lett. {\bf 59},
2582 (1987).
\bibitem{elastic} In our system anchoring at the free isotropic--nematic 
interface is planar, and the homeotropically--oriented nascent wetting nematic film 
will change to a linearly distorted configuration, with a faster growth rate (from
logarithmic to power law) on approaching coexistence; of course 
this does not invalidate our argument. See D. E. Sullivan and R. Lipowsky, 
Can. J. Chem.
{\bf 66}, 553 (1988); T. J. Sluckin and A. Poniewierski, Mol. Cryst. Liq.
Cryst. {\bf 179}, 349 (1990); F. N. Braun, T. J. Sluckin and E. Velasco, 
J. Phys.: Condens. Matter {\bf 8}, 2741 (1996).
\bibitem{Chiccoli} C. Chiccoli, P. Pasini, A. Sarlah, C. Zannoni, and S. Zumer,
Phys. Rev. E {\bf 67}, 050703 (2003).
\bibitem{Galabova} H. G. Galabova, N. Kothekar, and D. W. Allender, Liq.
Crys. {\bf 23}, 803 (1997).
\bibitem{Sarlah} A. Sarlah and S. Zumer, Phys. Rev. E {\bf 60}, 1821 (1999).
\bibitem{Zappone} B. Zappone, Ph. Richetti, R. Barberi, R. Bartolino, and 
H. T. Nguyen, Phys. Rev. E {\bf 71}, 041703 (2005).
\bibitem{Dani1} D. de las Heras, L. Mederos, and E. Velasco,
Phys. Rev. E {\bf 68}, 031709 (2003).
\bibitem{Dani2} D. de las Heras, E. Velasco, and L. Mederos,
J. Chem. Phys. {\bf 120}, 4949 (2004).
\bibitem{Doug} D. J. Cleaver and M. P. Allen, Mol. Phys. {\bf 80},
253 (1993).
\bibitem{Ising} F. Brochard--Wyart and P. G. de Gennes, C. R. Acad.
Sci. Paris II {\bf 297}, 223 (1983); A. O. Parry and R. Evans,
Phys. Rev. Lett. {\bf 64}, 439 (1990); M. R. Swift, A. L. Owczarek, 
and J. O. Indekeu, Europhys. Lett. {\bf 14}, 475 (1991); for more
recent references see e. g. Ref. \cite{Quintana}.
\bibitem{Inma1} I. Rodr\'{\i}guez--Ponce, J. M. Romero--Enrique, and L. F. 
Rull, J. Chem. Phys. {\bf 122}, 014903 (2005).
\bibitem{Inma} I. Rodr\'{\i}guez--Ponce, J. M. Romero--Enrique, and L. F. 
Rull, Phys. Rev. E {\bf 64}, 051704 (2001).
\bibitem{Quintana} J. Quintana and A. Robledo, Physica A {\bf 248}, 28 (1998).
\bibitem{Zap} The S phase may have been observed already 
in the recent surface--force--apparatus 
masurements of Zappone et al. \cite{Zappone}, who probed nematic films
under homeotropic/planar hybrid conditions. At very short film thicknesses
($<10$ nm) the force between the surfaces becomes strongly attractive,
a signature that, according to the authors, may point to a film
reconstruction into a step--like phase.
\end{thebibliography}
\end{document}